\documentclass[a4paper,12pt]{article}  
  
\usepackage{amsfonts,amssymb}  
\usepackage{theorem}  

\newcommand{\Sf}[1]{\tilde{#1}}

\def\G{\Gamma}    
   
\def\Sbv{S_{\mathrm{BV}}}

\def\dbfm{\hat \delta}

\begin{document}     

\rightline{YITP-SB-04-09}
\rightline{MPP-2004-36}
\rightline{March 23$^{rd}$, 2004}
\vskip 1.5 truecm

\Large
\bf

\centerline{The background field method and the}
\centerline{linearization problem for Poisson manifolds}

\normalsize
\rm

\vskip 1.5 truecm
\normalsize
\centerline{P.~A.~Grassi$^{~a,b,}$\footnote{~pgrassi@insti.physics.sunysb.edu},
            A.~Quadri$^{~c,}$\footnote{~quadri@mppmu.mpg.de}}
\vskip .7 truecm
\centerline{$^{(a)}$ 
{\it C.N. Yang Institute for Theoretical Physics,} }  
\centerline{\it State University of New York at Stony Brook,   
NY 11794-3840, USA}  
\vskip .3cm  
\centerline{$^{(b)}$ {\it Dipartimento di Scienze e Tecnologie Avanzate,}} 
\centerline{\it Universit\`a del Piemonte Orientale} 
\centerline{\it C.so Borsalino, 54,  Alessandria, 15100 Italy}  
\vskip .3cm  
\centerline{$^{(c)}$ {\it Max-Planck-Institut f\"ur Physik (Werner-Heisenberg-Institut)}}
\centerline{\it F\"ohringer Ring 6, D-80805 M\"unchen, Germany}

\vskip 0.7 truecm
\normalsize
\bf
\centerline{Abstract}

\rm
\begin{quotation}
The background field method (BFM) for the Poisson Sigma Model (PSM) is studied 
as an example of the application of the BFM technique
to open gauge algebras. The relationship with Seiberg-Witten
maps arising in non-commutative gauge theories is clarified.
It is shown  that 
the implementation of the BFM for the PSM in the Batalin-Vilkovisky formalism
is equivalent to
the solution of a generalized linearization problem 
(in the formal sense) for Poisson structures
 in the presence of gauge fields.
Sufficient conditions for the existence of a solution and a constructive
method to derive it are presented.
\end{quotation}

\newpage

\section{Introduction}\label{sec:intro}

The background field method (BFM) technique  
\cite{Abbott:1980hw,Abbott:zw}
is by now a well-established theoretical tool in the framework of gauge
and string theory.
The BFM relies on the definition of a suitable splitting
of the original quantum fields $\Phi^\alpha$ into a background classical $\hat \phi^\alpha$
and a new quantum part $\phi^\alpha$ (so that the path integral is now carried
out over the $\phi^\alpha$-variables), designed in such a way to derive
a new set of local Ward-Takahashi (WT) identities, which are linear in the 
quantum fields $\phi^\alpha$.

Provided a suitable choice of the gauge-fixing functional has been
performed, the BFM WT identities for the quantum effective action
$\G$ hold together  with the Slavnov-Taylor (ST) identities,
translating at the quantum level the invariance of the
gauge-fixed classical action under the BRST symmetry.

The non-linearity of the BRST transformations $s \phi^\alpha(x)$ of the quantum fields
requires the introduction of a suitable set of external sources $\phi_\alpha^*(x)$,
known as antifields 
\cite{zj,bv,Gomis:1994he}, in order to control the renormalization
of the local operators $s\phi^\alpha(x)$. This results in the bilinear form of
the ST identities (also known as master equation within the 
Batalin-Vilkovisky (BV) formalism \cite{bv,Gomis:1994he}) for the quantum vertex functional
$\G$:
\begin{eqnarray}
(\G, \G) = 0 \, , 
\label{intro_1}
\end{eqnarray}
where the BV bracket in the L.H.S. of the above equation is defined by
\begin{eqnarray}
(X,Y) = 
\int d^4x \sum_i \Big ( \frac{\delta_r X}{\delta \phi^\alpha}
                    \frac{\delta_l Y}{\delta \phi^*_\alpha} -
                    \frac{\delta_r X}{\delta \phi^*_\alpha}
                    \frac{\delta_l Y}{\delta \phi^\alpha} \Big ) \, .
\label{intro_2}
\end{eqnarray}
As a consequence of the bilinear form of the BV bracket, the ST
identities in eq.(\ref{intro_1}) relate 1-PI Green functions
at different orders in the loop expansion.
On the contrary, the background gauge invariance, being linear in the
quantum fields, yields WT identitites that 
connect 1-PI Green functions at the same order in perturbation theory. 

In order to control the interplay between the ST identities 
and the BFM WT identities the BV bracket
in eq.(\ref{intro_2}) has to be extended to the background fields,
under the prescription that the latter form BRST doublets together with 
the corresponding background ghosts.  
This in turn guarantees \cite{Barnich:2000zw,Quadri:2002nh,Piguet:er} that the
BRST cohomology of the underlying theory (and hence 
its physical content) is not affected by the introduction
of the background fields themselves.

Then the BFM equivalence theorem \cite{Abbott:zw,Becchi:1999ir,Ferrari:2000yp}
 states that the Green functions
of physical BRST invariant operators can be computed by starting
from the renormalized background gauge-invariant effective action,
fulfilling the extended ST identities, after dropping
the dependence on the quantum fields $\phi^\alpha$.
The (physical) connected functions are then obtained
by taking the Legendre transform w.r.t. the background
fields $\hat \phi^\alpha$, once a suitable gauge-fixing 
for the classical background fields is introduced 
\cite{Abbott:1980hw,Grassi:1995wr,Grassi:1997mc}.

This property, together with the linear relations between 1-PI
Green functions stemming from the BFM WT identities, allows for
appealing simplifications in the calculation of physical quantities,
which have been widely exploited in the literature \cite{Kluberg-Stern:rs}-
\cite{Bornsen:2002hh}.

The geometrical aspects of the BFM are best appreciated in more
general situations where the requirement that the BFM transformations
are linear in the quantum fields can actually be regarded as
the condition on the existence of a suitable set of coordinates
with a privileged r\^ole.

As an example, in the case of non-linear sigma models \cite{Howe:vm} 
the  gauge transformations are replaced by diffeomorphisms
of the target manifold $M$. 
The BFM is then implemented by performing a non-linear splitting
of the original quantum fields $\Phi^\alpha$, which can be thought of as
coordinates on the target manifold $M$, as a function of
 the background fields $\hat \phi^\alpha$ and of 
the tangent vectors $\xi^\alpha$, spanning
the tangent space $T_P M$ with $P$ the point belonging to $M$
of coordinates $\{ \hat \phi^\alpha\}$.
Being tangent vectors,
$\xi^\alpha$ transform linearly under the BFM symmetry.
In a more geometrical language, the implementation of the BFM is equivalent 
to the transition to the set of normal coordinates \cite{Honerkamp:1971sh,Alvarez-Gaume:hn} in a neighborhood $U$ of the point $P$.

This example also shows quite clearly
that the BFM expresses a local property
of the model at hand: it is not known if all possible
points $\{ \Phi^\alpha \}$ can be parameterized in terms of the
normal coordinates centered around 
the background configuration $P=\{ \hat \phi^\alpha \}$. On general grounds,
$U$ is a proper subset of $M$.

\medskip
The extension of the BFM to the case of open gauge algebras
displays in an even more surprising way the local nature of the 
BFM and its geometrical interpretation as a condition on the
existence of suitably generalized normal coordinates.

In \cite{Grassi:2003mv} a first attempt in this direction
was made by studying the implementation of the BFM in the
case of Euclidean N=2  Super Yang-Mills theory 
in the Wess-Zumino (WZ) gauge.
The BFM was meant to yield a linearized version of the
full set of symmetries of N=2 SYM, thus including 
N=2 supersymmetry. As is well-known, in the WZ gauge
the supersymmetry is non-linearly realized. Therefore the existence of a
background splitting, achieving a linearized version of supersymmetry
transformations at the background level, is a highly non-trivial point.
Indeed the fact that such a problem can be solved,
so that a BFM formulation of the full symmetry content of N=2 SYM
can be found, proves to be the BFM counterpart of the existence of 
a topological symmetry in N=2 SYM, originally discovered by Witten
in his seminal paper \cite{Witten:1988ze}.

In the case of open gauge algebras the condition that the
BFM splitting does not modify the cohomology of the model at hand
becomes the condition of the canonicity (w.r.t. the BV bracket)
of the background field redefinition.
For N=2 SYM the canonicity of the BFM splitting is best
proven by keeping the auxiliary fields appearing in the WZ gauge
and by carrying out a field redefinition only involving the fields
of the theory. This is automatically a canonical field redefinition.
However, in the auxiliary field representation the symmetry
algebra of N=2 SYM is a closed one. 

\medskip
In this paper we extend our analysis to 
the implementation of the BFM
for the Poisson Sigma Model  (PSM) 
\cite{Ikeda:1993fh,Schaller:1994es,Cattaneo:1999fm}. For a general
Poisson tensor the symmetry algebra of the PSM is an
open gauge algebra for which no equivalent auxiliary field
representation in terms of a closed one is known.
Hence the quantization of the PSM must be carried
out by exploiting the full Batalin-Vilkovisky formalism \cite{bv,Gomis:1994he},
with the antifields playing an essential r\^ole in this context. The 
application of the BV formalism in the context of PSM has been 
formulated in \cite{kummer}. In addition, the study of consistent 
deformation of the PSM along the lines of \cite{hb} has been performed 
in \cite{Izawa:1999ib}. 

Besides the technical challenges and the interesting mathematical 
properties of the PSM, we also have to mention the application in string 
theory. In a given limit of NS-NS background of the full-fledge string 
theory, the worldsheet sigma model is a Poisson sigma model 
(see for example \cite{BLN}). 

The results we found are simple enough that we can state them
now. 
The PSM contains in the bosonic sector a set of gauge fields
$\bar \eta_{i,\mu}$ and a set of matter fields $\bar X^i$, 
the latter parameterizing
the target Poisson manifold $M$ on which
the Poisson tensor $\alpha^{ij}(\bar X)$ is defined.
 To these fields one must add
the ghost fields $\bar \beta_i$ 
associated to the (open) gauge algebra of the PSM.
The implementation of the BFM for the PSM proves to be equivalent
to the construction of a set of coordinates where the 
Poisson tensor $\alpha^{ij}$ is linear (linearization problem of Poisson
structures).

Such a linearization problem can be first studied in the restricted
sector spanned by $(\bar X^i, \bar \beta_i)$ (dropping the
gauge fields $\eta_{i,\mu}$ for the moment). By using cohomological
methods one can prove that the existence of a map, achieving
the linearization of the Poisson structure, can be thought as the
construction of a suitable Seiberg-Witten map similar to the one
appearing in non-commutative gauge theories \cite{NC}-\cite{Jurco:2000ja}.
The relevant Wess-Zumino (WZ) consistency conditions are verified as a
consequence of the fact that $\alpha^{ij}$ is a Poisson tensor.
One can give a sufficient condition on the possibility to solve
the linearization problem. The linear term in $\alpha^{ij}(\bar X) =
{f^{ij}}_k \bar X^k + O(\bar X^2)$ 
defines a set of structure constants ${f^{ij}}_k$, associated to
a Lie algebra $\mathfrak{G}$.
If $\mathfrak{G}$ is semi-simple\footnote{For PSM with non-semisimple 
algebra see for example \cite{zucch}.},
 then the linearization problem can be solved
by a formal power series field redefinition.

This reproduces the result obtained in the mathematical literature
by Weinstein \cite{weinstein}.

\medskip
If one considers the same problem in the presence of the
gauge fields, one finds that no solution exists in the general case.
This is because the WZ consistency condition associated with
the Seiberg-Witten map for the gauge fields is not verified unless
the Poisson tensor $\alpha^{ij}(\bar X)$ is linear in the $\bar X$'s.
This reflects the fact that in the general case the gauge 
algebra of the PSM is open.

Therefore one needs to reformulate the linearization problem
in the BV scheme by taking into account the antifields.
This is best done in the superfield formalism developed in
\cite{Cattaneo:1999fm}. It turns out that the requirement of
canonicity of the BFM field redefinition w.r.t to the relevant
BV antibracket can be given a precise geometrical interpretation
in terms of the tensorial transformation properties of
the even superfield $\Sf{\bar X}^i$ and the odd superfield $\Sf{\bar \eta}_i$,
gathering the fields and the antifields of the PSM.

This condition allows us to prove that, under the assumption that 
the associated Lie algebra $\mathfrak{G}$ is semi-simple, the linearization
problem can be actually solved in the phase 
space spanned by the fields and the antifields of the model. 

We see that the implementation of the BFM program in the PSM 
is able to provide
a sound geometrical interpretation (w.r.t. the Poisson geometry of $M$)
of the BV prescription.
Special attention has to be paid to 
the condition of canonicity of the change of variables
in the full phase space of the theory. We remind that 
there is no solution of the linearization problem by a
field redefintion which does not involve the antifields.

\medskip
We would like to comment on the local nature of these results.
Under the assumption that 
the associated Lie algebra $\mathfrak{G}$ at a given point
$P$ on the manifold $M$ is semi-simple, 
we are able to show by using the BFM technique that the PSM is locally
equivalent to a quantum gauge theory with an ordinary Lie gauge algebra
$\mathfrak{G}$, provided quantization is carried out according to the standard
perturbative methods.
We warn against the fact that 
no further conclusion about the global behaviour of the PSM
can be drawn.\footnote{Non-perturbative treatment of PSM 
has been consider in \cite{Schaller:1994es,kummer, Hirshfeld:1999xm}.} 

For instance, it may happen that 
the Poisson tensor $\alpha^{ij}$ does not have the same rank
on the whole manifold $M$. 
As a consequence, $M$ splits into a set of submanifolds, 
each endowed with a constant-rank Poisson structure.
It is not expected that a given BFM splitting, which is valid
on one of these submanifolds, is able to control the behaviour of the fields
on the other submanifolds. Otherwise said, the BFM in the formulation
we are discussing here does not seem to provide access to global information
on the PSM.
This is a point which we think deserves to be best studied in the future.

It is also worthwhile to notice that some Poisson structures
of physical interest, like the quantum plane,
do not meet the condition that $\mathfrak{G}$ is semi-simple.
The question of whether a BFM formulation of these
theories exists is also an interesting problem to be clarified.

At the end, we mention the existence of graded PSM 
\cite{Grumiller:2002nm} and the application to dilationic supergravity. 
The application of the BFM to those models can be done 
by using a graded Seiberg-Witten map. 

\medskip
The paper is organized as follows.
In Sect.~\ref{sec:PSM} we give a short introduction on the PSM
and analyze its BV quantization procedure.
In Sect.~\ref{sec:SW} we approach the linearization problem
by means of cohomological methods close to those developed in
non-commutative gauge theories and show that the linearization
map can be regarded as a special type of Seiberg-Witten map.
We clarify the obstructions to the construction of the SW map
in the presence of gauge fields and point out that 
the extension of the phase space to the antifields is needed
in order to formulate properly the BFM problem for the PSM.
In Sect.~\ref{sec:superfield} we analyze the PSM in the superfield
formalism and argue that the requirement of canonicity (w.r.t.
the BV bracket) of the linearization transformation 
can be understood in geometrical terms as a condition on
the tensorial transformation properties of the 
superfields of the PSM.
This condition is powerful enough to show that the linearization problem
can be solved (provided that the associated Lie algebra
$\mathfrak{G}$ is semi-simple).
In Sect.~\ref{sec:BFM} we explicitly construct the BFM for the PSM
and comment on the BFM gauge-fixing condition suited for the model at hand.
Finally conclusions are presented in Sect.~\ref{sec:conclusions},
while the Appendices
contain some additional technical material.

\section{The Poisson Sigma Model}\label{sec:PSM}

\subsection{A brief review}  
  
We take over the notation of Ref.~\cite{Cattaneo:1999fm}.  
Moreover, we will denote by a bar the fields of the starting model, before  
the linearization procedure and the corresponding
background splitting are performed.
  
The Poisson Sigma Model (PSM) has two real bosonic   
fields $\bar X,\bar \eta$.   
$\bar X$ is a map from the disc $D=\{u\in {\bf R}^2,\, |u|\leq 1\}$ to  
the target manifold   
$M$ while $\bar \eta$ is a differential 1-form on $D$ taking values in  
the pull-back by $\bar X$ of the cotangent bundle of $M$,
i.e. a section of $\bar X^*(T^*M)\otimes T^*D$.  
We denote by $n$ the dimension of the manifold $M$.

We can introduce local coordinates in which $\bar X$ is
represented by $d$ functions $\bar X^i(u)$. The latter
parameterize the manifold $M$.
In this representation $\bar \eta$ is given by 
$d$ differential 1-forms $\bar \eta_i(u)=  
\bar \eta_{i,\mu}(u)du^\mu$.

The action of the Poisson Sigma Model is  
\begin{eqnarray}  
S[\bar X,\bar \eta]=\int_D\bar \eta_i(u)\wedge d\bar X^i(u)+{\frac12}\,\alpha^{ij}  
(\bar X(u))\bar \eta_i(u)\wedge\bar \eta_j(u).  
\label{PSM_action}
\end{eqnarray}
where the Poisson tensor $\alpha^{ij} (\bar X(u))$ satisfies 
the Jacobi identities 
\begin{eqnarray}  
\alpha^{ls}\partial_s \alpha^{ij} + \alpha^{js}\partial_s \alpha^{li}  
+ \alpha^{is}\partial_s \alpha^{jl} = 0 \, .
\label{poiss_e17}  
\end{eqnarray}  

The gauge fields $\bar \eta$ are supplemented with the   
 boundary condition that for  
$u\in \partial D$, $\bar \eta_i(u)$ vanishes on vectors tangent  
to $\partial D$.   
  
\medskip
The action of the PSM is invariant under
the following infinitesimal gauge transformations  
with infinitesimal parameters $\bar \beta_i$:  
\begin{eqnarray*}  
\delta_{\bar \beta} \bar X^i&=&\alpha^{ij}(\bar X)\bar \beta_j,\\  
\delta_{\bar \beta} \bar \eta_i&=&-d{\bar\beta}_i-  
\partial_i\alpha^{jk}(\bar X){\bar \eta}_j{\bar \beta}_k.  
\end{eqnarray*}  

We point out that in the general case 
the algebra of the gauge transformations closes only on-shell:
\begin{eqnarray}  
{}[\delta_{\bar \beta},\delta_{\bar \beta'}]\bar X^i&=  
&\delta_{\{\bar \beta,\bar \beta'\}}{\bar X}^i,\\  
{}[\delta_{\bar \beta},\delta_{\bar \beta'}]\bar \eta_i&=&  
\delta_{\{\bar \beta,\bar \beta'\}}{\bar \eta}_i  
-\partial_i\partial_k\alpha^{rs}{\bar \beta}_r{\bar \beta'}_s  
(d{\bar X}^k+\alpha^{kj}(\bar X){\bar \eta}_j).  
\label{psm1}
\end{eqnarray}  
Here $\{\bar \beta,\bar \beta'\}_i=-\partial_i\alpha^{jk}(\bar X)  
\bar \beta_j\bar \beta'_k$.
$d\bar X^k+\alpha^{kj}\bar \eta_j=0$  
is the equation of motion for $\bar \eta_k$
derived from the action $S$.  
In this calculation the Jacobi identity for the 
Poisson tensor $\alpha^{ij}$ plays a fundamental r\^ole. 

The equation of motion for $\bar \eta_k$ appears in the R.H.S. of
eq.(\ref{psm1}) multiplied by the second derivative of the Poisson
tensor. 
If $\alpha^{ij}(\bar X)$ is linear in the $\bar X$'s then 
the algebra of the gauge transformations in eq.(\ref{psm1})
closes off-shell.

In this case $M$ is the dual space to a Lie algebra ${\cal G}$ 
with Kirillov--Kostant Poisson structure. 
The Lie bracket of two linear functions $f,g\in {\cal G} = M^*$  
is just the Poisson bracket and is again a linear function on $M$.  
Then the classical action is best viewed as a function of   
a field $\bar X$ taking values in ${\cal G}^*$ and a connection $d+\bar\eta$  
on a trivial principal bundle on $D$.   
After an integration by parts, the action becomes the ``BF action''  
$S=\int_D \langle \bar X,F(\bar \eta)\rangle$   
where $F(\bar \eta)$ is the curvature of $d+\bar \eta$. 
In this case the gauge transformation is the usual gauge transformation  
(with gauge parameter $-\bar\beta$)  
of a connection and a field $\bar X$ in the coadjoint representation. 

Another special case is when $\alpha=0$.  
Then the action is invariant under translations  
of $\bar \eta$ by exact one-forms on $D$. 
On the other hand
if $\alpha^{ij}$ is an invertible matrix (symplectic case)
one can formally integrate 
over $\bar \eta$ to get the action $\int_D \bar X^*\omega$.
The latter 
is invariant under arbitrary translations $\bar X^i\mapsto  
\bar X^i+\xi^i$, with $\xi^i(u)=0$ on the boundary of $D$. 

\medskip
In the BRST formalism one then promotes the  
infinitesimal gauge parameter $\bar \beta_i$ to an anticommuting  
ghost field (vanishing on the boundary of the disc)  
and introduces the BRST operator $s$
such that
\begin{eqnarray}  
s \bar X^i=\alpha^{ij}(\bar X)\bar \beta_j\,, ~~~
s \bar \eta_i=-d\bar \beta_i-\partial_i\alpha^{kl}(\bar X)  
\bar \eta_k\bar \beta_l\,, ~~~ 
s \bar \beta_i=\frac12\,  
\partial_i\bar \alpha^{jk}(\bar X)\bar \beta_j\bar\beta_k.  
\label{eq:brst}
\end{eqnarray}  
Then $s$ is a differential on shell, i.e., it squares  
to zero modulo the equation of motion of $\bar \eta_k$. Indeed we get
$s^2\bar X^i=s^2\bar \beta_i=0$ and  
$s^2\bar\eta_i=  
-\frac12\,\partial_i\partial_k\alpha^{rs}\bar \beta_r\bar \beta_s  
(d\bar X^k+\alpha^{kj}(\bar X)\bar \eta_j).  
$  
One can assign a gradation, the ghost number, to the fields:  
$\mathrm{gh}(\bar X^i)=\mathrm{gh}(\bar \eta_i)=0$, $\mathrm{gh}(\bar \beta_i)=1$.  
The BRST operator has then ghost number one. 
Additionally  
there exists the gradation of the fields as differential forms  
on the disc, which will be denoted by deg: $\mathrm{deg}(\bar X^i)  
=\mathrm{deg}(\bar \beta_i)=0$, $\mathrm{deg}(\bar \eta_i)=1$.  
  
In the case $M={\cal G}^*$ of linear Poisson structures, the  
second derivatives of $\alpha$ vanish, and the BRST operator  
squares to zero.  

\medskip  
In order to implement the BFM in the PSM we first need to find,
if it exists, a suitable set of coordinates achieving the 
linearization of the Poisson  
tensor ${\alpha}^{ij}$.
We first observe that by Weinstein's splitting theorem \cite{weinstein}
locally around any point $P$ of a Poisson manifold $M$, spanned
by the coordinates $x^a$, $a=1,\dots,n$ and equipped with
the Poisson tensor $w^{ab}(x)$, it is possible to find
a canonical set of coordinates $(q^1,\dots,q^k,p^1,\dots,p^k,y^1,\dots,y^s)$
such that their Poisson brackets, defined by
\begin{eqnarray}
\{ F, G \} = w^{ab} \frac{\partial F}{\partial x^a}\frac{\partial G}{\partial x^b} \, ,
\label{psm4}
\end{eqnarray}
are
\begin{eqnarray}
&& \{ q^i, q^j \} = \{ p^i, p^j \} = \{ q^i, y^r \} = \{ p^i, y^r \} = 0 \, ,
\nonumber \\
&& \{ q^i, p^j \} = \delta^{ij} \, , ~~~~~ \{ y^r, y^t \} = c^{rt}(y)
\, ,
\label{psm5}
\end{eqnarray}
with $c^{rt}(0) =0$. The sector spanned by $(q^i, p^j)$ is associated
to a symplectic Poisson submanifold. We are only interested here
in the non-symplectic case (the 
sector described by the coordinates $(y^1, \dots, y^s)$ with Poisson
tensor ${c^{rt}}(y)$). 

Therefore we will restrict ourselves to a Poisson tensor $\alpha^{ij}(\bar X)$
vanishing at $\bar X=0$.
Moreover,  since we work within perturbative quantum field theory, we also
require that ${\alpha}^{ij}(\bar X)$ is a formal power series in $\bar X$. 
Since $\alpha^{ij}(0)=0$, $\alpha^{ij}(\bar X)$ starts
at least with terms linear in $\bar X$.
The linear terms in its expansion  
\begin{eqnarray}  
{\alpha}^{ij}(\bar X) = {f^{ij}}_k \bar X^k + \dots   
\label{psm2}  
\end{eqnarray}  
identify the structure constants ${f^{ij}}_k$ of a Lie algebra $\mathfrak{G}$ 
\cite{weinstein} whose generators $T^i$ fulfill  
\begin{eqnarray}  
[T^i,T^j] = - i {f^{ij}}_k T^k \, .  
\label{poiss_e2}  
\end{eqnarray}  
The corresponding Poisson tensor $\alpha_0^{ij}(X) \equiv {f^{ij}}_k X^k$ 
gives rise to a linear Poisson  structure whose associated BRST differential is  
\begin{eqnarray}  
s X^i = {f^{ij}}_k X^k \beta_j \, , ~~~~ 
s \eta_i = -d \beta_i - {f^{jk}}_i \eta_j \beta_k \, , ~~~~ 
s \beta_i = \frac{1}{2} {f^{jk}}_i \beta_j \beta_k \, .  
\label{poiss_e1}  
\end{eqnarray}  
%
%

From the BRST point of view the linearization problem can then be stated 
as follows: find a suitable change of coordinates  
\begin{eqnarray}  
\bar X^i = \bar X^i (X, \eta) \, , ~~~~ 
\bar \beta_i = \bar \beta_i(\beta,X,\eta) \, , ~~~~ 
\bar \eta_i = \bar \eta_i(X,\eta) \, ,  
\label{poiss_e3}  
\end{eqnarray}  
fulfilling the following initial conditions  
\begin{eqnarray}  
&& \bar X^i = X^i + \mbox{terms of order at least 2 in } X \, , ~~~~ 
\nonumber \\
&& 
\bar \beta_i = \beta_i + \mbox{terms of order at least 1 in } X \, ,  ~~~~
\nonumber \\
&&
\bar \eta_i = \eta_i + \mbox{terms of order at least 1 in } X \, ,  
\label{poiss_e4}  
\end{eqnarray}  
and such that the following transformation rules
hold
\begin{eqnarray}  
&& s \bar X^i = \alpha^{ij}(\bar X) \bar \beta_j \, , \nonumber \\  
&& s \bar \eta_i = - d \bar \beta_i - \partial_i \alpha^{jk} (\bar X)   
   \bar \eta_j \bar \beta_k \, , \nonumber \\  
&& s \bar \beta_i = \frac{1}{2} \partial_i \alpha^{jk} (\bar X)   
   \bar \beta_j \bar \beta_k \, ,  
\label{poiss_e5}  
\end{eqnarray}  
where $\alpha^{ij}(\bar X)$ is the original Poisson tensor 
and the new variables $X^i, \beta_i$ and $\eta_i$
transform as in eq.(\ref{poiss_e1}) under the BRST differential $s$.
  
By taking into account the ghost number of the fields
we see that $\bar \beta_i$ must be linear in $\beta_i$.
  
Once a solution to the linearization problem is found, 
a linear splitting for the new variables  
$X^i=\hat X^i + Q^i$, $\eta_i= \hat \eta_i + \xi_i$ 
can be performed to implement the BFM.

\medskip
Several comments are in order here. The linearization problem,
as stated above, can be considered in the restricted space
of the variables $(\bar X^i,\bar \beta_i$), dropping the
gauge fields $\bar \eta_i$. 
Such a restriction is well studied in the mathematical
literature and goes under the name of linearization
of Poisson brackets (see \cite{fernandes} for a recent review).
Specifically, one Taylor-expands the Poisson tensor
$c^{rt}(y)$ in eq.(\ref{psm5}) around zero
and separates its linear part as
follows:
\begin{eqnarray}
c^{rt}(y) = {c^{rt}}_v y^v + g^{rt}(y)
\label{psm8}
\end{eqnarray}
with ${c^{rt}}_v = \frac{\partial c^{rt}(y)}{\partial y^v}(0)$,
$g^{rt}(y) = c^{rt}(y) - {c^{rt}}_v y^v$.
The linearization problem can be stated in the following way:
are there new coordinates where the functions $g^{rt}(y)$
identically vanish, yielding a bracket which is linear
in these new coordinates?

If $M$ is an analytic (resp. $C^\infty$-) manifold 
equipped with the Poisson tensor $c^{rt}(y)$ and the linearization
problem can be solved by an analytic (resp. $C^\infty$-) change of variables,
we say that such a Poisson structure is analytically 
(resp. $C^\infty$-) linearizable.
If the linearization can be established by a local
formal power series, one speaks instead of a formally linearizable
Poisson structure.
Since in the framework of Quantum Field Theory we only deal
with Poisson tensors which are formal power series we will
restrict ourselves to the study of formal linearization.

We notice that we will actually consider an extension of the
linearization problem studied so far in the mathematical context,
namely we will deal with the full set of fields of the PSM
including the gauge fields $\eta_i$.
This in turn will force us to pose the problem and
to look for its solution in the more general framework provided
by the Batalin-Vilkovisky formalism, as we explain in the next
Section.

\subsection{BV formulation of the PSM}\label{sec:BV}

Since the BRST differential $s$ in eq.(\ref{eq:brst}) squares to
zero modulo the equations of motion, the Batalin-Vilkovisky (BV)
formalism \cite{bv,Gomis:1994he} is needed in order to construct the
classical action fulfilling the relevant master equation.
Terms quadratic in the antifields will appear as a consequence
of the on-shell nilpotency of $s$.

We follow the discussion given in \cite{Cattaneo:1999fm}.
For each of the fields $\phi_\alpha$ of the model
(with $\phi_\alpha$ standing for
$\bar X^i, \bar \beta_i$ and $\bar\eta_{i,\mu}$) the corresponding
antifield, denoted by $\phi^*_\alpha$, is introduced. 
The assignment of ghost and degree number to the
antifields is given in the following table.

\begin{center}
\begin{tabular}{|c|c|c|}
\hline
$\phi^*_\alpha$ & gh($\phi^*_\alpha$) & deg($\phi^*_\alpha$)  \cr
\hline
$\bar X^*_i$ & -1 & 0 \cr
\hline
$\bar \beta^{i*}$ & -2 & 0 \cr
\hline
$\bar \eta^{i*}_\mu$ & -1 & 1 \cr
\hline
\end{tabular}
\end{center}
The ghost number is assigned according to the relation
\begin{eqnarray}
\textrm{gh}(\phi^*_\alpha) = -\textrm{gh}(\phi^\alpha)-1 \, .
\label{gh_number}
\end{eqnarray}
The classical action is required to have zero ghost number.
The statistics of the antifield $\phi^*_\alpha$ is opposite
to that of $\phi^\alpha$, namely
\begin{eqnarray}
\epsilon(\phi^*_\alpha) = \epsilon(\phi^\alpha) + 1 ~~ (\textrm{mod } 2) \, .
\label{stat}
\end{eqnarray}
We have $\epsilon(\bar X^i)=\epsilon(\bar\eta_{i,\mu})=0$, 
$\epsilon(\bar \beta_i)=1$.

The BV bracket is then defined according to
\begin{eqnarray}
(X,Y) & = &
 \sum_\alpha \int _D dv(u) \, \Big ( <\frac{\delta_r X}{\delta \phi^\alpha},
                    \frac{\delta_l Y}{\delta \phi^*_\alpha}>
                  - <\frac{\delta_r X}{\delta \phi^*_\alpha},
                    \frac{\delta_l Y}{\delta \phi^\alpha}> \Big ) \, .
\label{pr1}
\end{eqnarray}
In the above equation $dv(u)$ is the volume
element $\sqrt{g}~ du^1 du^2$ associated with a Riemannian
metric $g_{\mu\nu}$ defined on $D$ and $<~,~>$ is the corresponding induced
scalar product on the exterior algebra of the cotangent space at $u$.
$\frac{\delta_l}{\delta \phi^\alpha}$ denotes the left
derivative w.r.t to $\phi^\alpha$, 
$\frac{\delta_r}{\delta \phi^\alpha}$ the right derivative and
analogously for $\phi^*_\alpha$.

%
%

In particular we obtain the following fundamental  BV brackets for the
variables of the PSM:
\begin{eqnarray}
(\bar X^i, \bar X^*_j) = \delta^i_j \, , ~~~~~
(\bar \beta^i, \bar \beta^*_j) = \delta^i_j \, , ~~~~
(\bar \eta^{i\mu}, \bar \eta^*_{j\nu}) = \delta^i_j \delta^\mu_\nu \, .
\label{pr3}
\end{eqnarray}

It is also useful to introduce the Hodge dual $\phi^+_\alpha$
of the antifields $\phi^*_\alpha$ according to 
\begin{eqnarray}
\phi^*_\alpha = * \phi^+_\alpha \, .
\label{pr4}
\end{eqnarray}
The Hodge dual fulfills $< \alpha, \beta > dv(u) = \alpha \wedge * \beta.$

In terms of $\phi^+_\alpha$ the BV bracket in eq.(\ref{pr1})
can be expressed without any reference to the Riemannian metric as
follows:
\begin{eqnarray}
(X,Y) = \sum_\alpha \int_D \Big (
\frac{\delta_r X}{\delta \phi^\alpha} \wedge 
\frac{\delta_l Y}{\delta \phi^+_\alpha} - (-1)^{\textrm{deg}(\phi_\alpha)}
\frac{\delta_r X}{\delta \phi^+_\alpha} \wedge
\frac{\delta_l Y}{\delta \phi^\alpha} \Big ) \, .
\label{bv_1}
\end{eqnarray}
According to the BV procedure one needs to find an extension $\Sbv$
of the classical action $S$ in eq.(\ref{PSM_action})
\begin{eqnarray}
\Sbv = S + S^{(1)} + S^{(2)} + \dots \, , 
\label{bv_2}
\end{eqnarray}
where $S^{(j)}$ is of degree $j$ in the number of the antifields,
satisfying the master equation
\begin{eqnarray}
(\Sbv, \Sbv)=0 \, .
\label{bv_3}
\end{eqnarray}
$S^{(1)}$ is obtained by coupling the antifield $\phi^*_\alpha$ with
the BRST variation of the corresponding field $\phi^\alpha$.
For the PSM $S^{(1)}$ is given by (in the Hodge dual notation for
the antifields)
\begin{eqnarray}
S^{(1)} = \int_D \bar X^+_i s \bar X^i + 
\bar \eta^{+i} \wedge s \bar \eta_i - \bar \beta^{+i} s \bar \beta_i \, .
\label{bv_4}
\end{eqnarray}
For the PSM the sum in the R.H.S. of eq.(\ref{bv_2}) ends at the second 
order, which is given by
\begin{eqnarray}
S^{(2)} = - \frac{1}{4} \int_D \bar \eta^{+i} \wedge \bar \eta^{+j} \partial_i 
\partial_j \alpha^{kl}(\bar X) \bar \beta_k \bar \beta_l \, .
\label{bv_5}
\end{eqnarray}
The full action satisfying the master equation is therefore
\begin{eqnarray}
\Sbv & = & \int_D\bar \eta_i \wedge d\bar X^i +{\frac12}\,\alpha^{ij}  
(\bar X)\bar \eta_i \wedge\bar \eta_j
\nonumber \\
& & ~~~ + \bar X^+_i \alpha^{ij}(\bar X) \bar \beta_j + 
\bar \eta^{+i} \wedge  ( - d \bar \beta_i - \partial_i \alpha^{jk} (\bar X)   
   \bar \eta_j \bar \beta_k)  
\nonumber \\
& & ~~~
-  \frac{1}{2} \bar \beta^{+i} \partial_i \alpha^{jk} (\bar X)   
   \bar \beta_j \bar \beta_k 
- \frac{1}{4} \bar \eta^{+i} \wedge \bar \eta^{+j} \partial_i 
\partial_j \alpha^{kl}(\bar X) \bar \beta_k \bar \beta_l \, .
\label{bv_6}
\end{eqnarray}
The differential naturally associated with the BV bracket, defined by
\begin{eqnarray}
\delta Y = (\Sbv, Y) \, ,
\label{bv_7}
\end{eqnarray}
is off-shell nilpotent, as a consequence of the master equation
in eq.(\ref{bv_3}).
Explicitly it acts as follows on the fields
\begin{eqnarray}
&& \delta \bar X^i = \alpha^{ij}(\bar X) \bar \beta_j \, , \nonumber \\
&& \delta \bar \beta_i = \frac{1}{2} \partial_i \alpha^{kl}(\bar X)\bar \beta_k \bar \beta_l
\, , \nonumber \\
&& \delta \bar \eta_i = -d \bar \beta_i - \partial_i \alpha^{kl}(\bar X)
\bar \eta_k \bar \beta_l - \frac{1}{2} \partial_i \partial j 
\alpha^{kl}(\bar X) \bar \eta^{+j} \bar \beta_k \bar \beta_l 
\label{bv_8}
\end{eqnarray}
and the antifields of the model
\begin{eqnarray}
&& \delta \bar X^+_i = d \bar \eta_i + \partial_i \alpha^{kl}(\bar X)
\bar  X^+_k \bar \beta_l - \partial_i \partial_j \alpha^{kl}(\bar X)
\bar \eta^{+j} \wedge \bar \eta_k \bar \beta_l 
+ \frac{1}{2} \partial_i \alpha^{kl}(\bar X) \bar \eta_k \wedge
\bar \eta_l \, , \nonumber \\
&& ~~~~~ - \frac{1}{4} \partial_i \partial_j \partial_p
\alpha^{kl}(\bar X) \bar \eta^{+j} \wedge \bar \eta^{+p}
\bar \beta_k \bar \beta_l 
- \frac{1}{2} \partial_i \partial_j \alpha^{kl}(\bar X)
\bar \beta^{+j} \bar \beta_k \bar \beta_l \, ,
\nonumber \\
&& \delta \bar \beta^{+i} = -d \bar \eta^{+i} 
- \alpha^{ij}(\bar X) \bar X^+_j + \frac{1}{2} \partial_k \partial_l
\alpha^{ij}(\bar X) \bar \eta^{+k} \wedge \bar \eta^{+l} \bar \beta_j
\nonumber \\
&& ~~~~~ + \partial_k \alpha^{ij}(\bar X) \bar \eta^{+k} \wedge \bar \eta_j
         + \partial_k \alpha^{ij}(\bar X) \bar \beta^{+k} \bar \beta_j \, , 
\nonumber \\
&& \delta \bar \eta^{+i} = -d \bar X^i - \alpha^{ij}(\bar X) \bar \eta_j
- \partial_k \alpha^{ij}(\bar X) \bar \eta^{+k} \bar \beta_j \, .
\label{bv_9}
\end{eqnarray}
The action of $\delta$ on the fields is the same as the one of $s$
with the exception of the antifield-dependent term in the 
last line of eq.(\ref{bv_8}).
The latter disappears if $\alpha^{ij}$ is linear in $\bar X$.
In such a case the BV action
$\Sbv$ in eq.(\ref{bv_6}) only contains terms linear in the antifields.
This corresponds to the well-known situation of an ordinary gauge theory
whose BRST differential $s$ is nilpotent. The master equation
in eq.(\ref{bv_3}) is then 
the Slavnov-Taylor identity for the model at hand.

\medskip
We remark that $\Sbv$ in eq.(\ref{bv_6}) is not gauge-fixed.
The inclusion of a gauge-fixing can be performed in the
standard way within the BV formalism \cite{Gomis:1994he} 
by extending the field
content to a non-minimal sector including the antighost fields
and then by performing a canonical transformation generated
by a suitable gauge-fixing functional $\Psi$. This has been
explicitly carried out in Ref.~\cite{Cattaneo:1999fm}.
In Sect.~\ref{sec:BFM} we will provide a short
discussion of the gauge-fixing procedure
which is relevant for the application of the background field method to the PSM.

\medskip
We also wish to comment on the way the BV formalism has been introduced.
In this section the use of the BV technique was motivated by the
fact that the relevant BRST differential $s$ squares to zero only on-shell.
Under these circumstances the BV method provides a almost systematic 
procedure to generate an extended action fulfilling
the BV master equation. Nevertheless the geometry of the extended
set of variables with the antifields remains somehow obscure.
In Sect.~\ref{sec:superfield} we will show that such a structure
naturally arises within the linearization problem in the presence
of the gauge fields $\bar \eta_{i,\mu}$.
Moreover, the action of the BV differential $\delta$
in eqs.(\ref{bv_8}) and (\ref{bv_9}) is precisely the one
required by the condition that the linearization procedure is carried
out (at least locally)
via a canonical (w.r.t to the BV bracket in eq.(\ref{bv_1})) 
field redefinition.
This in turn provides a better insight into the geometrical
structure of the PSM.

\section{The Seiberg-Witten map for the PSM}\label{sec:SW}

In this section we analyze the linearization problem as stated
in eqs.(\ref{poiss_e1})-(\ref{poiss_e5}) by means of cohomological 
methods close to those applied in the context of 
non-commutative gauge theories \cite{NC}-\cite{Barnich:2001mc}
in order to prove the existence of a Seiberg-Witten (SW) map
\cite{Jurco:2001rq}-\cite{Jurco:2000ja}.
In this formulation the antifields are not involved.

\subsection{The linearization problem for $\bar X^i,\bar \beta_i$}
\label{sec:lin}

We first discuss the set of equations given by the first and the third
of eqs.(\ref{poiss_e5}). Moreover, we make the ansatz that 
these equations can be solved by a field redefinition only
involving $X^i,\beta_i$, thus excluding any mixing with the
gauge fields $\eta_{i,\mu}$. Therefore we restrict ourselves
to field redefinitions of the type
\begin{eqnarray}
\bar X^i= \bar X^i(X) \, , ~~~~ \bar \beta_i = \bar \beta_i(\beta,X) \, .
\label{sw1}
\end{eqnarray}
By the ghost number $\bar \beta_i$ must be linear in $\beta_i$.

\medskip
There is  a natural grading induced by the counting operator 
for the $X$, given by  
\begin{eqnarray}  
{\cal N} = \int d^4x \, X^i \frac{\delta}{\delta X^i} \, .  
\label{poiss_e6}  
\end{eqnarray}  
Notice that $s$ in eq.(\ref{poiss_e1}) is of order zero with respect  
to this grading. We will solve the first of eqs.(\ref{poiss_e5}) 
order by order in the grading induced by ${\cal N}$.  
  
At the lowest order a solution compliant with the boundary
conditions in eq.(\ref{poiss_e4}) is given by  
\begin{eqnarray}  
\bar X^i = X^i \, , ~~~~ \bar \beta_i = \beta_i \, .
\label{poiss_e7}  
\end{eqnarray}  
We now assume that the first and the third of eqs.(\ref{poiss_e5}) 
are fulfilled up to order $n-1$:  
\begin{eqnarray}  
&& s \bar X^{i \, (l)} =   
\left [ \alpha^{ij} (\bar X) \bar \beta_j   
\right ]^{(l)} \, , 
\nonumber \\
&& s \bar \beta^{(l)}_i = \frac{1}{2}   
\left [ \partial_i \alpha^{jk}(\bar X) \bar \beta_j \bar \beta_k  
\right ]^{(l)}  
 ~~~~~~~~~~~~ l=0,1,\dots,n-1 \, .   
\label{poiss_e8}  
\end{eqnarray}  
We denote the order in the grading induced by ${\cal N}$ by a superscript  
in parentheses, so that $\bar X^{i \, (l)}$  is the component  
of $\bar X^i$ of order $l$ in the $X$.  
At the $n$-th order we have 
\begin{eqnarray}  
s \bar X^{i \, (n)}  & = &  \sum_{p+q=n} \alpha^{ij \, (p)}   
\bar \beta^{(q)}_j \, ,  \nonumber \\
s \bar \beta_i^{(n)} & = & \frac{1}{2} \sum_{p+q+r=n}   
\left [ \partial_i \alpha^{jk}\right ]^{(p)} \bar \beta_j^{(q)}   
\bar \beta_k^{(r)} \, , 
\label{poiss_e9}  
\end{eqnarray}  
where $ \alpha^{ij \, (p)}$ is a shorthand notation for  
\begin{eqnarray}  
 \alpha^{ij \, (p)} = [ \alpha^{ij} (\bar X(X)) ]^{(p)}   
\label{poiss_e10}  
\end{eqnarray}  
and similarly for $\bar \beta_j^{(q)}$.  
  
We put in evidence the $\bar X^{i \, (n)}$-dependent contribution from the  
R.H.S. of the first of eqs.(\ref{poiss_e9}) and write  
\begin{eqnarray}  
s \bar X^{i \, (n)} & = &  
\sum_{p+q=n, \, p \neq 0, \, q \neq 0}   
\alpha^{ij \, (p)} \bar \beta_j^{(q)} + \alpha^{{ij} \, (n)} \beta_j   
\nonumber \\  
& = & \sum_{p+q=n, \, p \neq 0, \, q \neq 0}   
\alpha^{ij \, (p)} \bar \beta^{(q)}_j +   
{f^{ij}}_k \bar X^{k \, (n)} \beta_j  \nonumber \\  
& & + \left [ \alpha^{{ij} \, (n)} \beta_j -  
{f^{ij}}_k \bar X^{k \, (n)} \beta_j \right ] \, .  
\label{poiss_e11}  
\end{eqnarray}  
The terms in the bracket is $\bar X^{i \,(n)}$-independent. Therefore we obtain  
\begin{eqnarray}  
s \bar X^{i \, (n)}  
- {f^{ij}}_k\bar X^{k \, (n)} \beta_j & = &  
 \sum_{p+q=n, \, p \neq 0, \, q \neq 0}   
\alpha^{ij \, (p)} \bar \beta^{(q)}_j \nonumber \\  
& &   
 +  \left [ \alpha^{{ij} \, (n)} \beta_j -  
{f^{ij}}_k \bar X^{k \, (n)} \beta_j \right ] \, .  
\label{poiss_e12}  
\end{eqnarray}  
In a similar fashion we get for the second 
of eqs.(\ref{poiss_e9})
\begin{eqnarray}
s \bar \beta_i^{(n)} & = & \frac{1}{2} \sum_{p+q+r=n}   
\left [ \partial_i \alpha^{jk} \right ]^{(p)} \bar \beta_j^{(q)}   
\bar \beta_k^{(r)} \nonumber \\  
& = &    
\frac{1}{2}   
\sum_{\scriptstyle p+q+r=n,  \scriptstyle p\neq 0, q\neq 0,r\neq 0 }  
\left [ \partial_i \alpha^{jk} \right ]^{(p)} \bar \beta_j^{(q)}   
\bar \beta_k^{(r)} \nonumber \\   
& & + \frac{1}{2}   
      \left [ \partial_i \alpha^{jk} \right ]^{(n)} \beta_j \beta_k  
\nonumber \\  
& & + \frac{1}{2} \left [ \partial_i \alpha^{jk} \right ]^{(0)}   
   \bar \beta^{(n)}_j \beta_k  
 + \frac{1}{2} \left [ \partial_i \alpha^{jk} \right ]^{(0)}   
   \beta_j \bar \beta^{(n)}_k     \nonumber \\  
& = &    
\frac{1}{2}   
\sum_{\scriptstyle p+q+r=n, \scriptstyle p\neq 0, q\neq 0,r\neq 0}  
\left [ \partial_i \alpha^{jk} \right ]^{(p)} \bar \beta_j^{(q)}   
\bar \beta_k^{(r)} \nonumber \\   
& & + \frac{1}{2}   
      \left [ \partial_i \alpha^{jk} \right ]^{(n)} \beta_j \beta_k  
+ {f^{jk}}_i \bar \beta^{(n)}_j \beta_k \, .  
\label{poiss_e21}  
\end{eqnarray}  
The above equation can be rewritten as  
\begin{eqnarray}  
&&   
\!\!\!\!\!\!\!\!\!\!\!\!\! 
s \bar \beta_i^{(n)} +{f^{jk}}_i \bar \beta_k^{(n)} \beta_j   
 \nonumber \\   
&& = \frac{1}{2}   
\sum_{\begin{array}{c}
\scriptstyle p+q+r=n,\cr
\scriptstyle p\neq 0, q\neq 0,r\neq 0 
\end{array} }  
\left [ \partial_i \alpha^{jk} \right ]^{(p)} \bar \beta_j^{(q)}   
\bar \beta_k^{(r)}   
+ \frac{1}{2}   
  \left [ \partial_i \alpha^{jk} \right ]^{(n)} \beta_j \beta_k   
\label{poiss_e22}  
\end{eqnarray} 
The L.H.S. of the eqs.(\ref{poiss_e12}) and (\ref{poiss_e22})
contains the following coboundary operator  
(expressed in terms of Lie algebra commutators):  
\begin{eqnarray}  
&& \Delta = s - i [\beta, \cdot] ~~~~ \mbox{on even quantities}  
\nonumber \\  
&& \Delta = s - i \{ \beta, \cdot \} ~~~~ \mbox{on odd quantities}   
\label{poiss_e13}  
\end{eqnarray}  
This is a well-known operator in non-commutative gauge field theories.  
It was introduced in \cite{Brace:2001fj} 
in order to formulate the WZ consistency  
condition for the Seiberg-Witten map. 
By construction $\Delta$ is nilpotent. 

We can rewrite eqs.(\ref{poiss_e12}) and (\ref{poiss_e22}) as follows:
\begin{eqnarray}
\Delta \bar X^{i \, (n)} = {\cal B}^{i (n)} \, , ~~~~
\Delta \bar \beta_i^{(n)} = {\cal A}_i^{(n)} \, ,
\label{cohom_eq}
\end{eqnarray}
with
\begin{eqnarray}  
{\cal B}^{i (n)} \equiv  \sum_{p+q=n, \, p \neq 0, \, q \neq 0}   
\alpha^{ij \, (p)} \bar \beta^{(q)}_j   
 +  \left [ \alpha^{{ij} \, (n)} \beta_j -  
{f^{ij}}_k \bar X^{k \, (n)} \beta_j \right ] \,   
\label{poiss_e14}  
\end{eqnarray}  
and 
\begin{eqnarray}  
{\cal A}^{(n)}_i \equiv \frac{1}{2}   
\sum_{\begin{array}{c} \scriptstyle p+q+r=n, \cr \scriptstyle  
 p\neq 0, q\neq 0,r\neq 0 \end{array}}  
\left [ \partial_i \alpha^{jk}(\bar X) \right ]^{(p)} \bar \beta_j^{(q)}   
\bar \beta_k^{(r)}   
+ \frac{1}{2}   
  \left [ \partial_i \alpha^{jk}(\bar X) \right ]^{(n)} \beta_j \beta_k   
\, .
\label{poiss_e25}  
\end{eqnarray}  
Notice that ${\cal B}^{i (n)}$, ${\cal A}^{(n)}_i$ 
only depend on  known lower orders terms. 
One can regard the linearization problem as the construction 
of a suitable SW map fulfilling eqs.(\ref{cohom_eq}).

\medskip
Therefore by applying it to the two eqs.(\ref{cohom_eq}) we find that,
if a solution exists, the following Wess-Zumino consistency conditions
must be satisfied:  
\begin{eqnarray}  
\Delta {\cal B}^{i (n)} = 0 \, , ~~~~ \Delta {\cal A}^{(n)}_i = 0 \, .  
\label{poiss_e15}  
\end{eqnarray}  
It can be easily checked by a straightforward computation that the
condition
\begin{eqnarray}
\Delta {\cal B}^{i (n)} = 0
\label{poiss_e16}
\end{eqnarray}
is verified. One needs to use the recursive assumption in 
eq.(\ref{poiss_e8}), the Jacobi identity for the
structure constants ${f^{ij}}_k$ as well as the Jacobi identity
for the Poisson tensor $\alpha^{ij}$  (\ref{poiss_e17}). 

The check that also the second Wess-Zumino consistency condition
\begin{eqnarray}
\Delta {\cal A}^{(n)}_i = 0
\label{poiss_e18}
\end{eqnarray}
is verified is a little bit more involved. The details of the calculation
are presented in Appendix \ref{appA}. Again it turns out that
eq.(\ref{poiss_e18}) is true thanks to the recursive
assumption in eq.(\ref{poiss_e8}) and to the 
fact that $\alpha^{ij}$ is a Poisson tensor.

\medskip
We would like to make a comment here. In non-commutative
gauge theories  an
analogous WZ consistency condition naturally arises for the SW map.
In that case it happens that the WZ consistency condition
is verified as a consequence of the associativity of the $\star$-product
\cite{NC}.
In the present case the WZ consistency conditions are instead intimately
related to the Poisson geometry.
The formal similarity between the two constructions is rather suggestive
and might lead to a deeper understanding of the common geometrical properties of the SW map \cite{kontsevich-cattaneo}.

\medskip
Having established the WZ consistency conditions in eq.(\ref{poiss_e15}),
the existence of the linearization map and hence of a solution 
to eqs.(\ref{cohom_eq}) boils down to prove that the operator $\Delta$
admits a homotopy ${\cal K}$ such that
\begin{eqnarray}
\{ \Delta, {\cal K} \} = \iota 
\label{hom_1}
\end{eqnarray}
where $\iota$ is the identity in the positive ghost number sector.

If such an operator exists, then the most general solution to the equation
\begin{eqnarray}
\Delta \Upsilon^{(n)} = \Lambda^{(n)}
\label{hom_2}
\end{eqnarray}
(where $\Upsilon^{(n)}$ stands for $\bar X^{i,(n)}$, $\bar \beta^{(n)}_i$
and $\Lambda^{(n)}$ respectively for ${\cal B}^{i (n)}$, ${\cal A}_i^{(n)}$)
is
\begin{eqnarray}
\Upsilon^{(n)} = {\cal K} \Lambda^{(n)} + \tau^{(n)}
\label{hom_3}
\end{eqnarray}
with $\tau^{(n)}$ an arbitrary element of the kernel of $\Delta$
with appropriate ghost number
parameterizing the ambiguities in the solution.

The question of the existence of a homotopy for $\Delta$ 
has been addressed in \cite{Brace:2001fj}. On general grounds
one can prove \cite{weinstein,fernandes} that
such an operator exists provided that the structure constants
${f^{ij}}_k$ are those of a semisimple Lie algebra.
Under this assumption the analysis presented here reproduces
one of the results of Weinstein's linearization theorem  \cite{weinstein},
stating that a Poisson structure, whose associated
Lie algebra\footnote{The Lie algebra associated to a Poisson tensor
$\alpha^{ij}(\bar X)$ vanishing at the origin is defined
by the structure constants ${f^{ij}}_k$ given by the coefficients
of the linear
term in $\alpha^{ij}$:
$$\alpha^{ij}(\bar X) = {f^{ij}}_k \bar X^k + \dots \, .$$
}
is semisimple, is formally non-degenerate,
i.e. there exists a change of variables
\begin{eqnarray}
\bar X^i = f^i(X)
\label{cov}
\end{eqnarray}
(in the sense of formal power series) such that in the 
variables $X$ the Poisson tensor is linear.

The construction given in the present Section constitutes
an explicit recursive cohomological method for obtaining the functions
$f^i(X)$. We notice that the use of cohomological techniques
in the spirit of the SW map relies on the introduction of
the ghost fields $\bar \beta_i$ (which do not appear in
Weinstein's proof). This is a natural feature from the 
field theoretic-point of view, since these ghost fields are
naturally associated to the gauge symmetries of the PSM.

\subsection{Inclusion of the gauge fields}

We now discuss the changes that happen if one wants to solve
the full system of eqs.(\ref{poiss_e5}), by taking into account also
the gauge fields $\eta_{i,\mu}$.
We relax the ansatz in eq.(\ref{sw1}) and consider
more general field redefinitions like those
in eq.(\ref{poiss_e3}).

The analysis of the first and third of eqs.(\ref{poiss_e5}) proceed
as before, on the contrary there is a problem with the second of
eq.(\ref{poiss_e5}). 
By following the same path as in Sect.~\ref{sec:lin}  we assume that
such an equation is fulfilled up to order $n-1$ in the number
of $X$'s and project it at the $n$-th order:
\begin{eqnarray}  
s \bar \eta_i^{(n)} & = & - d \bar \beta_i^{(n)}   
                     - \left [ \partial_i \alpha^{jk}(\bar X) \bar \eta_j  
                               \bar \beta_k \right ]^{(n)}   
\nonumber \\  
& = & - d \bar \beta_i^{(n)} -  
\sum_{\scriptstyle p+q+r=n, \scriptstyle  
 q\neq n}     
\left ( \partial_i \alpha^{jk}(\bar X) \right )^{(p)}   
\bar \eta_j^{(q)} \bar \beta_k^{(r)}   
- {f^{jk}}_i \bar \eta_j^{(n)} \beta_k \, .   
\label{poiss_e36}  
\end{eqnarray}  
The above equation can be rewritten as  
\begin{eqnarray}  
s \bar \eta_i^{(n)} - {f^{jk}}_i \bar \eta_k^{(n)} \beta_j = {\cal E}^{(n)}_i  
\label{poiss_e37}  
\end{eqnarray}  
where  
\begin{eqnarray}  
{\cal E}^{(n)}_i \equiv - d \bar \beta_i^{(n)} -  
\sum_{\scriptstyle p+q+r=n,  \scriptstyle  
 q\neq n}     
\left ( \partial_i \alpha^{jk}(\bar X) \right )^{(p)}   
\bar \eta_j^{(q)} \bar \beta_k^{(r)} \, .  
\label{poiss_e38}  
\end{eqnarray}  
Equivalently we can write  
\begin{eqnarray}  
\Delta \bar \eta_i^{(n)} = {\cal E}_i^{(n)} \, .  
\label{poiss_e39}  
\end{eqnarray}  
where $\Delta$ is defined according to eq.(\ref{poiss_e13}).  
In view of the nilpotency of $\Delta$ we derive 
a new WZ consistency condition
\begin{eqnarray}
\Delta {\cal E}_i^{(n)}  = 0 \, .
\label{poiss_e40}
\end{eqnarray}
By evaluating the L.H.S. of the above equation
we get however that it does not vanish:
\begin{eqnarray}  
\Delta {\cal E}^{(n)}_i = -\frac{1}{2} \beta_m \beta_k  
\left [ \partial_i \partial_l \alpha^{mk}   
\left ( d \bar X^l + \alpha^{lj}(\bar X) \bar \eta_j \right ) \right ]^{(n)} \, .  
\label{poiss_e41}  
\end{eqnarray}  
From eq.(\ref{poiss_e41}) 
we see that in the general case the Wess-Zumino consistency condition  
is verified only if the fields are restricted to the subspace  
where the equation  
of motion for $\bar \eta_l$ holds true, i.e. the constraint  
\begin{eqnarray}  
d \bar X^l + \alpha^{lj}(\bar X) \bar \eta_j = 0 \,   
\label{poiss_e42}  
\end{eqnarray}  
is imposed. 
Moreover, from eq.(\ref{poiss_e41}) the WZ consistency condition
is also verified in the case of a linear Poisson tensor
(without requiring the validity of the equation of motion
for $\bar \eta_l$).

This result is rather remarkable. It means that for arbitrary Poisson
structures
a solution to the linearization problem involving only the fields 
of the model does not exist.  
This is a direct consequence of the fact that 
the algebra of the PSM is open. 
As we will show, the antifields need to be taken into account 
in order to formulate properly the linearization problem
in the presence of the gauge fields $\eta_{i,\mu}$.  
This in turn will shed new light on the geometrical
meaning of the BV construction of the PSM.

\section{The superfield formalism}\label{sec:superfield}

In this section we show that the 
 BV formalism naturally appears in the 
linearization of the Poisson structure
in the presence of the gauge fields $\eta_{i,\mu}$.

The BV construction of the PSM described in Sect.~\ref{sec:BV}
can be recasted in the superfield formalism \cite{Cattaneo:1999fm}.
For that purpose one introduces in addition to the even coordinates
$u^1,u^2$ on $D$ two odd (anticommuting) coordinates 
$\theta^1,\theta^2$.
The fields and the antifields 
of the PSM are gathered into even superfields $\Sf{\bar X^i}$ 
\begin{eqnarray}
\Sf{\bar X^i} = \bar X^i + \theta^\mu \bar \eta^{+i}_\mu 
- \frac{1}{2} \theta^\mu \theta^\nu \bar \beta^{+i}_{\mu\nu}
\label{sf1}
\end{eqnarray}
and odd superfields $\Sf{\bar \eta_i}$
\begin{eqnarray}
\Sf{\bar \eta_i} = \bar \beta_i + \theta^\mu \bar \eta_{i,\mu} 
+ \frac{1}{2} \theta^\mu \theta^\nu \bar X^+_{i,\mu\nu} \, .
\label{sf2}
\end{eqnarray}
We define $D = \theta^\mu \partial/\partial u^\mu$. The BV differential
$\delta$ in eq.(\ref{bv_7}) acts as follows on the 
superfields $\Sf{\bar X^i}$, $\Sf{\bar \eta_i}$:
\begin{eqnarray}
&& \delta \Sf{\bar X^i} = D \Sf{\bar X^i} + \alpha^{ij}(\Sf{\bar X}) \Sf{\bar \eta_j} \, , \nonumber\\
&& \delta \Sf{\bar \eta_i} = D \Sf{\bar \eta_i} 
+ \frac{1}{2} \partial_i \alpha^{jk}(\Sf{\bar X}) \Sf{\bar \eta_j}
\Sf{\bar \eta_k} \, .
\label{sf3}
\end{eqnarray}
By the Jacobi identity in eq.(\ref{poiss_e17}) obeyed by $\alpha^{ij}$ 
we find $\delta^2=0$.
The action on the components of the superfields can be obtained by
taking into account the expansion in eq.(\ref{sf1}) and (\ref{sf2})
and by projecting out the $\delta$-variations 
in eq.(\ref{sf3}) on the relevant powers of $\theta^\mu$. 
The transformation rules of the components fields and antifields
of course reproduce those given in eqs.(\ref{bv_8})
and (\ref{bv_9}).

The BV action is obtained by integrating the two-form part of $L$
\begin{eqnarray}
L^{(2)} = \int d^2 \theta ~ L
\label{sf4}
\end{eqnarray}
with 
\begin{eqnarray}
L = \Sf{\bar \eta_i} D \Sf{\bar X^i} + \frac{1}{2} \alpha^{ij}(\Sf{\bar X})
\Sf{\bar \eta_i} \Sf{\bar \eta_j} \, ,
\label{sf5}
\end{eqnarray}
so that
\begin{eqnarray}
\Sbv = \int_D L^{(2)} \, .
\label{sf6}
\end{eqnarray}
It is easy to prove that 
\begin{eqnarray}
\delta \Sbv = 0 \, ,
\label{sf7}
\end{eqnarray}
or equivalently that $\Sbv$ fulfills the master equation. This is because
\begin{eqnarray}
\delta L = D(\Sf{\bar \eta_i} D \Sf{\bar X^i}) \, ,
\label{sf8}
\end{eqnarray}
and hence $\delta L^{(2)}$ is the differential of a one-form vanishing
along the boundary.

\subsection{Superfield linearization transformation for the PSM}

We now wish to consider the linearization problem for the PSM within
the superfield formalism. 

First we notice that there exists a bracket in the superfield
space naturally generated by the BV bracket in eq.(\ref{bv_1}):
\begin{eqnarray}
\{ X, Y \} = \int d^2 \theta \,
\sum_\alpha \int _D 
\Big ( \frac{\delta_r X}{\delta \phi^\alpha}\wedge
                    \frac{\delta_l Y}{\delta \phi^+_\alpha}
                  - (-1)^{\textrm{deg}(\phi_\alpha)}
                    \frac{\delta_r X}{\delta \phi^+_\alpha}\wedge
                    \frac{\delta_l Y}{\delta \phi^\alpha} \Big ) \, .
\label{sf9}
\end{eqnarray}
%

With respect to the $\{, \cdot, \cdot \}$-bracket 
$\Sf{\bar X^i}$ and $\Sf{\bar \eta_i}$ are
conjugate variables, namely
\begin{eqnarray}
\{ \Sf{\bar X^i}, \Sf{\bar \eta_j} \} = \delta^i_j \, .
\label{sf10}
\end{eqnarray}
We now discuss the linearization problem for the PSM.
We impose the requirement 
that the linearization transformation will preserve
the bracket in eq.(\ref{sf10}), i.e. that is is canonical
w.r.t. the $\{ \cdot,\cdot \}$-bracket.

From the analysis of Sect.~\ref{sec:lin} we know that,
under the assumption that the associated Lie algebra is semi-simple,
there exists (at least one) set of functions $f^i$ in eq.(\ref{cov})
\begin{eqnarray}
\bar X^i = f^i(X)
\label{sf11}
\end{eqnarray}
solving the linearization problem in the restricted
$(X,\beta)$-space.

We now choose one set of such functions $f^i(X)$ and lift them
to the superfield space, i.e. we consider the transformation
\begin{eqnarray}
\Sf{\bar X^i} = f^i (\Sf X) \, .
\label{sf12}
\end{eqnarray}
In the above equation $\Sf{X^i}$ is the even superfield associated
with the new variables in which the Poisson tensor is linear:
\begin{eqnarray}
\Sf{X^i} = X^i + \theta^\mu \eta^{+i}_\mu 
- \frac{1}{2} \theta^\mu \theta^\nu \beta^{+i}_{\mu\nu} \, .
\label{sf13}
\end{eqnarray}
Analogously we define 
\begin{eqnarray}
\Sf{\eta_i} = \beta_i + \theta^\mu \eta_{i,\mu} 
+ \frac{1}{2} \theta^\mu \theta^\nu X^+_{i,\mu\nu} \, ,
\label{sf14}
\end{eqnarray}
where $\Sf{\eta_i}$ is canonically conjugated via
the $\{\cdot,\cdot\}$-bracket to $\Sf{X_i}$.

The extension of the transformation to the odd superfields
$\Sf{\bar \eta_i}$ is carried out under the requirement that
the whole map preserves the bracket in eq.(\ref{sf10})
(canonicity requirement).

This can be done by using the following generating
functional\footnote{This is a canonical transformation of the $q-P$-type: 
definining the fundamental bracket as $\{ q,p \}=1$ then the 
generating functional $S(q,P)$ gives rise
to the following canonical transformation w.r.t. the $\{,\}$-bracket:
\begin{eqnarray*}
p = \frac{\partial S}{\partial q}, ~~~~ Q = \frac{\partial S}{\partial P} \, ,
\end{eqnarray*}
where $(Q,P)$ are the new variables.}
\begin{eqnarray}
\Psi (\Sf{X}, \Sf{\bar \eta}) = f^i (\Sf{X}) \Sf{\bar \eta}_i \, .
\label{pr18}
\end{eqnarray}
The corresponding canonical superfield transformation is given by
\begin{eqnarray}
&& \Sf{\eta_j} = \frac{\partial f^i}{\partial \Sf{X^j}}  \Sf {\bar \eta_i} \, , \nonumber \\
&& \Sf{\bar X^i} = f^i(\Sf{X}) \, .
\label{pr19}
\end{eqnarray}
The first of the above equations is understood to be inverted in order to give
$\Sf{\bar \eta}_i$ as a function of $\Sf{\eta}, \Sf{X}$.
Notice that this is possible (in the sense of formal power series)
since by the boundary condition in eq.(\ref{poiss_e4}) we have
\begin{eqnarray}
\frac{\partial f^i}{\partial \Sf{X^j}} = \delta^i_j + O(\Sf{X}^2) \, .
\label{can1}
\end{eqnarray}

One can give a geometrical interpretation of eq.(\ref{pr19}). It states that
$\Sf{\bar \eta_i}$ transforms like a covector under the transformation generated by the maps $f^i$. By making use of this observation and of the
fact that $\alpha^{ij}$ is a Poisson tensor we now show that
the linearization problem is solved by the map in eq.(\ref{pr19}).
This clarifies the geometrical interpretation of the requirement
of canonicity within the superfield approach to the BV formalism.

For notational convenience we rewrite the superfield
redefinition in eq.(\ref{pr19}) as follows:
\begin{eqnarray}
&& \Sf{X}^i = \Sf{X}^i(\Sf{\bar X}) \, , \nonumber \cr
&& \Sf{\eta}_i = \frac{\partial \Sf{\bar X}^j}{\partial \Sf{ X}^i} 
\Sf{\bar \eta}_j \, .
\label{pr20}
\end{eqnarray}
The map $\Sf{X}^i = \Sf{X}^i(\Sf{\bar X})$ has been constructed 
in such a way that
in the $\Sf{X}$-variables the Poisson tensor is linear. Therefore
\begin{eqnarray}
\alpha^{ij}(\Sf{\bar X}(\Sf{X})) & = &  
                              \frac{\partial \Sf{\bar X}^i}
                                   {\partial \Sf{X}^p}
                              \frac{\partial \Sf{\bar X}^j}
                                   {\partial \Sf{X}^q}
	         	      \alpha^{pq}(\Sf{X}) \nonumber \\
                       & = &  \frac{\partial \Sf{\bar X}^i}
                                   {\partial \Sf{X}^p}
                              \frac{\partial \Sf{\bar X}^j}
                                   {\partial \Sf{X}^q}
                              {f^{pq}}_l \Sf{X}^l \, ,
\label{pr21}
\end{eqnarray}
since $\alpha^{ij}$ is a tensor of rank $(2,0)$.
\medskip
We now prove that the action of the BV differential $\delta$ on
the new variables $\Sf{X}^i, \Sf{\eta}_i$ is generated by the
linearized Poisson tensor which appears in the R.H.S. of eq.(\ref{pr21}).

\medskip
 We first consider the action of $\delta$ on $\hat X^i$. One has
\begin{eqnarray}
\delta \Sf{X}^i & = & \frac{\partial \Sf{X}^i}{\partial \Sf{\bar X}^j} 
                      \delta \Sf{\bar X}^j 
                  = \frac{\partial \Sf{X}^i}{\partial \Sf{\bar X}^j} 
                    ( D \Sf{\bar X}^j + \alpha^{jk}(\Sf{\bar X}) 
                        \Sf{\bar \eta}_k ) \nonumber \\
                & = & D \Sf{X}^i +  
                      \frac{\partial \Sf{X}^i}{\partial \Sf{\bar X}^j} 
                      \alpha^{jk}(\Sf{\bar X})
                      \frac{\partial \Sf{X}^r}{\partial \Sf{\bar X}^k} 
                      \Sf{\eta}_r
                    \nonumber \\
                & = &  D \Sf{X}^i + {f^{ir}}_l \Sf{X}^l \Sf{\eta}_r \, ,
\label{pr22}
\end{eqnarray}
which corresponds to the action of $\delta$ determined by the linearized
Poisson tensor $\alpha^{ij}(\Sf{X}) =  {f^{ij}}_k \Sf{X}^l$.

The analysis of the action of $\delta$ on $\Sf{\eta}_i$ is
a little bit more involved. The details of the computations
are given in Appendix \ref{appB}.

The final result is
\begin{eqnarray}
\delta \Sf{\eta}_i = D \Sf{\eta}_i +  \frac{1}{2} {f^{ab}}_i \Sf{\eta}_a \Sf{\eta}_b \, .
\label{fin_1}
\end{eqnarray}

From the above analysis 
we conclude that the linearization of the Poisson structure
in the presence of the gauge fields has been achieved by 
the transformation in eq.(\ref{pr20}) in the full 
superspace spanned by $\Sf{X}^i$ and
$\Sf{\eta}_i$. 
The introduction of the antifields (and hence of the superfields)
is essential in order to achieve such a result.
The canonicity of the superfield redefinition 
is guaranteed by the conservation of the bracket in
eq.(\ref{sf10}). The geometrical interpretation of eq.(\ref{pr20})
shows that the requirement of canonicity is precisely the condition
needed in order to achieve linearization in the unbarred supervariables,
as a consequence of the tensorial nature of $\alpha^{ij}$.

\medskip
We wish to comment on the locality property of the linearization
field redefinition for the PSM.
From the above analysis it turns out that 
the linearization procedure holds true in a suitable neighborhood of 
any point of the Poisson manifold where the Poisson structure
is linearizable (in the formal sense). The determination of the maximal
region $U$ where the parameterization of the Poisson manifold $M$
by means of the linearized variables is valid is a problem
which is beyond the cohomological techniques used to 
solve the linearization problem locally.
As an example, if the rank of the Poisson tensor is not constant
on $M$ one of course expects that $U$ will prove to be a proper subset of $M$. 

While showing that for linearizable Poisson tensors
one can always choose locally a suitable parameterization,
reducing the problem to that of an ordinary off-shell
closed Lie gauge algebra, the linearization technique
in its present form is nevertheless not developed enough in order to
analyze global features of the PSM quantization.

\section{The BFM for the PSM}\label{sec:BFM}

In this section we discuss how the bacground field method
(BFM) can be implemented in the PSM by making use of the
linearized superfield variables $(\Sf{X}^i, \Sf{\eta}_i)$.

The $L$-functional in eq.(\ref{sf5}) is given in terms
of the new variables by
\begin{eqnarray}
L = \Sf{\eta}_i D \Sf{X}^i + \frac{1}{2} {f^{ij}}_k \Sf{X}^k \Sf{\eta}_i \Sf{\eta}_j \, .
\label{bfm1}
\end{eqnarray}
When expressed as a function of the component fields
the corresponding BV action 
\begin{eqnarray}
\Sbv = \int_D L^{(2)} 
\label{bfm2}
\end{eqnarray}
reads
\begin{eqnarray}
\Sbv & = & \int_D \, \eta_i \wedge dX^i + \frac{1}{2} {f^{ij}}_k X^k \eta_i \wedge \eta_j \nonumber \\
     &   & ~~~~ + X^+_i {f^{ij}}_k X^k \beta_j - \eta^{+i} \wedge (d \beta_i + 
{f^{kl}}_i \eta_k \beta_l) \nonumber \\
&& ~~~~ -\frac{1}{2} \beta^{+i} {f^{jk}}_i \beta_j \beta_k \, .
\label{bfm3}
\end{eqnarray}
In $\Sbv$ all quadratic terms depending on the antifields have disappeared.
In order to implement the BFM for the PSM in the representation
given by eq.(\ref{bfm3}) we split linearly the gauge and matter fields
as follows
\begin{eqnarray}
\eta_{i,\mu} = \hat \eta_{i,\mu} + \xi_{i,\mu} \, , ~~~~
X^i = \hat X^i + Q^i \, ,
\label{bfm4}
\end{eqnarray}
where $\hat \eta_{i,\mu}$ are the background gauge fields,
$\hat X^i$ the background matter fields, $\xi_{i,\mu}$
the quantum gauge fields and $Q^i$ the quantum
matter fields.
Then the action in eq.(\ref{bfm3}) is invariant under the following
background transformation of parameters $\omega_i$:
\begin{eqnarray}
&& \dbfm \hat \eta_i = - d \omega_i - {f^{kl}}_i \hat \eta_k \omega_l \, , 
\nonumber \\
&& \dbfm \Phi^i = {f^{ij}}_k \Phi^k \omega_j \, , ~~~~~
   \dbfm \varphi_i = -{f^{kl}}_i \varphi_k \omega_l \, , 
\label{bfm5}
\end{eqnarray}
with $\Phi^i = \{ Q^i, \hat X^i, \beta^{+i} \}$, $\varphi_i = \{ \xi_i, \beta_i, \eta^+_i, X^+_i \}$.
The $\dbfm$-transformations of the fields are linear
in the quantized fields of the theory.
The linear splitting in eq.(\ref{bfm4}) is a consequence of the
linearization of the Poisson structure achieved by the superfield
redefinition analyzed in Sect.~\ref{sec:superfield}.

The BRST differential is extended in such a way to
include the background fields as BRST doublets:
\begin{eqnarray}
&& s \hat X^i = \Omega_X^i \, , ~~~~~~ s \Omega_X^i = 0 \, ,
\nonumber \\
&& s \hat \eta_{i,\mu} = \Omega_{i,\mu} \, , ~~~~ s \Omega_{i, \mu} = 0 \, ,
\label{bfm5_bis}
\end{eqnarray}
where $\Omega_X^i$ and $\Omega_{i,\mu}$ are the background ghosts
for the matter and the gauge fields respectively.
The BRST transformations of $\xi_{i,\mu}$ and $Q^i$ are determined by the
condition that the BRST transformation of the original fields
$X^i, \eta_{i,\mu}$ is preserved:
\begin{eqnarray}
&& s \xi_i = - d \beta_i - {f^{kl}}_i (\hat \eta_k + \xi_k) \beta_l - \Omega_i \, , \nonumber \\
&& s Q^i = {f^{ij}}_k (\hat X^k + Q^k) \beta_j - \Omega_X^i \, .
\label{bfm5_ter}
\end{eqnarray}

\medskip
In view of the fact that the gauge transformations for the PSM
are in the new unbarred variables those associated to an ordinary Lie algebra,
the BFM-gauge-fixing procedure can be carried out according to the
standard methods developed for Yang-Mills theory 
\cite{Abbott:1980hw,Grassi:1995wr}.
For that purpose we introduce the antighost fields $\gamma^i$
on $D$ and the corresponding Nakanishi-Lautrup multiplier fields
$\lambda^i$, together with their antifields $\gamma^+_i, \lambda^+_i$.
The ghost number of $\gamma^i$ is $-1$, that of $\lambda^i$
is zero.
The boundary condition for $\lambda^i$ is $\lambda^i(u)=0$, $u\in \partial D$,
while $\gamma^i$ is constant on the boundary.

The BRST differential acts of $\gamma^i,\lambda^i$ as follows:
\begin{eqnarray}
s \gamma^i = \lambda^i \, , ~~~~ s \lambda^i = 0 \, .
\label{bfm5_1}
\end{eqnarray}
The following BFM-gauge-fixing term is added to the classical action:
\begin{eqnarray}
S_{\textrm{gf}} = - \int_D \,  * d \gamma^i \wedge s ~(d \xi_i + {f^{kl}}_i 
\hat \eta_k \xi_l) + \lambda^i (d * \xi_i + {f^{kl}}_i \hat \eta_k * \xi_l)
\, .
\label{bfm6}
\end{eqnarray}
The BFM prescription is encoded in the last term of the above equation,
where the ordinary choice for the gauge-fixing function
$F_i =  d * \eta_i$ is 
replaced by its covariantized form with respect to the
background gauge field $\hat \eta_k$:
$$F_i = d * \xi_i + {f^{kl}}_i \hat \eta_k * \xi_l \, . $$

The full action $\G^{(0)}$, defined by
\begin{eqnarray}
\G^{(0)} = \Sbv + S_{\textrm{gf}} - \int_D \lambda^i \gamma_i^+ \, , 
\label{bfm7}
\end{eqnarray}
obeys the BV master equation and in addition is invariant under
the BFM differential $\dbfm$, 
with the following transformation rules for
$\gamma^i, \lambda^i$ and $\gamma_i^+, \lambda_i^+$:
\begin{eqnarray}
&& \dbfm \gamma^i = {f^{ij}}_k \gamma^k \omega_j \, , ~~~~~
\dbfm \gamma^+_i = -{f^{kl}}_i \gamma^+_k \omega_l \, , \nonumber \\
&& \dbfm \lambda^i = {f^{ij}}_k \lambda^k \omega_j \, , ~~~~~
\dbfm \lambda^+_i = -{f^{kl}}_i \lambda^+_k \omega_l \, .
\label{bfm8}
\end{eqnarray}
This construction implements the BFM for the PSM associated 
to a linearizable Poisson structure.

\section{Conclusions and Perspectives}\label{sec:conclusions}

The extension of the BFM to the case of open gauge algebras
has been carried out on the example of the Poisson Sigma Model.
For this model no equivalent representation in terms
of auxiliary fields, leading to a closed gauge algebra,
is known. Hence the full BV formalism is needed to handle
the PSM.

The request of linearity of the background transformation
singles out a privileged set of local coordinates,
in which the background gauge transformations of the model acquire
a simple form. For non-linear sigma models the BFM quantization
implies the transition to normal coordinates, for the PSM - whenever
this is possible, i.e. the Poisson structure
is (formally) linearizable - to the linearization coordinates $X^i$.

We have pointed out that the map solving the linearization problem
for the PSM is a Seiberg-Witten map of the same kind of those
appearing in the
framework of non-commutative gauge theories. This might suggest
a deeper understanding of the relationship between the PSM
and non-commutative geometry.

We have proven a no-go theorem, stating that in the general
case in the presence of gauge fields no change of variables, 
only involving the fields
of the theory, can lead to a solution of the linearization problem.
This happens because there is an obstruction to the fulfillment of the
associated Wess-Zumino consistency condition, which is equal to zero
only modulo the equations of motion for the gauge fields.

We find that  the antifields play an essential r\^ole in order
to formulate consistently and solve the linearization problem for the PSM
in the BV formalism.

We have clarified the 
r\^ole of the fundamental properties
of the BFM implementation, namely the linearity 
in the quantum fields of the background transformations and
the canonicity of the BFM splitting change of variables,
in the general context provided by the BV formalism and the related
superfield formulation for the PSM.
As far as canonicity is concerned, 
we have shown that the requirement that the BFM splitting
does not modify the cohomology of the model at hand
can be understood in pure geometrical terms
as a condition on the tensorial transformation properties
of the superfields of the model.
The latter property in turn allows to solve the linearization
problem for the PSM, under the condition that the
associated Lie algebra is semi-simple. 
The BFM can be finally implemented as a consequence of the solution
of the linearization problem for the relevant Poisson structure.

We wish to emphasize that the BFM construction for the PSM
is a local one. Indeed in the general case not all field
configurations can be covered by using the linearization coordinates,
for instance when the rank of the Poisson tensor is not constant
on the Poisson manifold $M$.

This suggest that a more refined quantization technique is
required in order to get a control of the global properties
of the PSM.

Another expect to be consider is the following: the relative 
simplicity of PSM (2d conformal field theory) permits the computation 
of the quantum correction to the SW map extending the relation beyond 
the classical level. As in \cite{Ooguri} it should be possible to show that 
the properties discussed so far in the present paper can be 
extended at the quantum level. The BFM is the fundamental 
tool required in order to perform such computations.

\section*{Acknowledgments}

One of us (A.Q.) would like to thank A.~Cattaneo for 
useful discussions. R. Stora and G. Barnich are acknowledged 
for suggestions and comments. The research of P.A.G. is partially funded 
by NSF GRANT PHY-0098527 and University of Piemonte Orientale. 
P.A.G. and A.Q. thank the Theory Division at CERN where 
this work was started and was completed. 

\appendix

\section{WZ consistency condition for the ghost field $\bar \beta_i$}
\label{appA}

We wish to verify eq.(\ref{poiss_e18}).  
First we compute  
\begin{eqnarray}  
&& \!\!\!\!\!\!\!\!\!\!\! s \Big ( \frac{1}{2}    
\sum_{\scriptstyle p+q+r=n, \scriptstyle  
 p\neq 0, q\neq 0,r\neq 0 }  
 \left [ \partial_i \alpha^{jk}(\bar X) \right ]^{(p)} \bar \beta_j^{(q)}   
\bar \beta_k^{(r)}  \Big )  = \nonumber \\  
&& ~~~~    
\frac{1}{2}   
\sum_{\begin{array}{c} \scriptstyle p+q+r=n, \cr \scriptstyle  
 p\neq 0, q\neq 0,r\neq 0 \end{array}}  
\Big \{  
 ( \partial_l \partial_i \alpha^{jk}(\bar X) s \bar X^l )^{(p)}  
 \bar \beta_j^{(q)} \bar \beta_k^{(r)} +   
 (\partial_i \alpha^{jk}(\bar X))^{(p)} s \bar \beta_j^{(q)}   
 \bar \beta_k^{(r)}  \nonumber \\  
&& ~~~~~~~~~~~~~~~~~~~~~~~  
- (\partial_i \alpha^{jk}(\bar X))^{(p)} \bar \beta_j^{(q)} s \bar \beta_k^{(r)}  
\Big \} \nonumber \\  
&&   
= \frac{1}{2} \Big \{  
\Big ( \partial_l \partial_i \alpha^{jk} \alpha^{lm}   
+ \frac{1}{2} \partial_i \alpha^{lk} \partial_l \alpha^{mj}   
- \frac{1}{2} \partial_i \alpha^{ml} \partial_l \alpha^{jk}  
\Big ) \bar \beta_m \bar \beta_j \bar \beta_k \Big \}^{(n)}  
\nonumber \\  
&&   
~~  - \frac{1}{2} \left ( \partial_l \partial_i \alpha^{jk} s \bar X^l  
   \right )^{(n)} \beta_j \beta_k - (\partial_i \alpha^{jk})^{(n)}   
   s \beta_j \beta_k \nonumber \\  
&&  
~~  
   - {f^{jk}}_i s \bar \beta^{(n)}_j \beta_k - {f^{jk}}_i s\beta_j \bar \beta_k^{(n)} \, ,
\label{poiss_e26}  
\end{eqnarray}  
where we have used the recursive assumption in eq.(\ref{poiss_e8}).

Since $\alpha^{ij}$ is a Poisson tensor, it fulfills the following  
equation  
\begin{eqnarray}  
\alpha^{ml} \partial_l \alpha^{kj} + \alpha^{jl}\partial_l \alpha^{mk}   
+ \alpha^{kl} \partial_l \alpha^{jm} = 0   
\label{poiss_e27}  
\end{eqnarray}  
and hence also  
\begin{eqnarray}  
\partial_i \left (   
\alpha^{ml} \partial_l \alpha^{kj} + \alpha^{jl}\partial_l \alpha^{mk}   
+ \alpha^{kl} \partial_l \alpha^{jm}  
\right ) = 0 \,   
\label{poiss_e28}  
\end{eqnarray}  
or  
\begin{eqnarray}  
\alpha^{ml} \partial_i \partial_l \alpha^{kj} + \mbox{cyclic } (m,k,j) =  
- \left ( \partial_i \alpha^{ml} \partial_l \alpha^{kj} +   
\mbox{cyclic } (m,k,j) \right ) \, .  
\label{poiss_e29}  
\end{eqnarray}  
We focus on the part between curly brackets in eq.(\ref{poiss_e26}),  
which we rewrite by taking into account antisymmetrization with respect  
to $m,j,k$ (denoted by square brackets) and eq.(\ref{poiss_e29}):  
\begin{eqnarray}  
&&  \partial_l \partial_i \alpha^{[jk} \alpha^{lm]}   
+ \frac{1}{2} \partial_i \alpha^{l[k} \partial_l \alpha^{mj]}   
- \frac{1}{2} \partial_i \alpha^{[ml} \partial_l \alpha^{jk]}   
\nonumber \\  
&& ~~~~~ = - \frac{1}{2} \partial_i \alpha^{[ml}\partial_l \alpha^{kj]}   
   - \frac{1}{2} \partial_i \alpha^{[ml}\partial_l \alpha^{kj]}   
\nonumber \\  
&& ~~~~~~~~ + \frac{1}{2} \partial_i \alpha^{l[k} \partial_l \alpha^{mj]}   
- \frac{1}{2} \partial_i \alpha^{[ml} \partial_l \alpha^{jk]} = 0  
\label{poiss_e30}  
\end{eqnarray}  
by noticing that   
\begin{eqnarray}  
 - \frac{1}{2} \partial_i \alpha^{[ml}\partial_l \alpha^{kj]}  =  
 + \frac{1}{2} \partial_i \alpha^{[lm}\partial_l \alpha^{kj]}  =  
 + \frac{1}{2} \partial_i \alpha^{[lk}\partial_l \alpha^{jm]} \, .  
\label{poiss_e31}   
\end{eqnarray}  
Hence we can rewrite eq.(\ref{poiss_e26}) as  
\begin{eqnarray}  
&& \!\!\!\!\!\!\!\!\!\!\!\!\!\!\!\!\!\!\!\!\!\!\! s \Big ( \frac{1}{2}    
\sum_{\scriptstyle p+q+r=n, \scriptstyle  
 p\neq 0, q\neq 0,r\neq 0 }  
 \left [ \partial_i \alpha^{jk}(\bar X) \right ]^{(p)} \bar \beta_j^{(q)}   
\bar \beta_k^{(r)}  \Big )  = \nonumber \\  
&&  
~~  - \frac{1}{2} \left ( \partial_l \partial_i \alpha^{jk} s \bar X^l  
   \right )^{(n)} \beta_j \beta_k - (\partial_i \alpha^{jk})^{(n)}   
   s \beta_j \beta_k \nonumber \\  
&&  
~~  
   - {f^{jk}}_i s \bar \beta^{(n)}_j \beta_k - {f^{jk}}_i s\beta_j \bar \beta_k^{(n)} \, .  
\label{poiss_e32}  
\end{eqnarray}  
We now compute  
\begin{eqnarray}  
&& s \left ( \frac{1}{2}   
  \left [ \partial_i \alpha^{jk}(\bar X) \right ]^{(n)} \beta_j \beta_k   
\right )   \nonumber \\  
&& ~~~~~~~~~~~~~~~  = \frac{1}{2}  
\left [ \partial_l \partial_i \alpha^{jk} s \bar X^l \right ]^{(n)}  
\beta_j \beta_k + \left [ \partial_i \alpha^{jk}(\bar X) \right ]^{(n)}  
 s \beta_j \beta_k \, .  
\label{poiss_e33}   
\end{eqnarray}  
By combining eqs.(\ref{poiss_e32}) and (\ref{poiss_e33}) we get  
\begin{eqnarray}  
s {\cal A}^{(n)}_i =  - {f^{jk}}_i s \bar \beta^{(n)}_j \beta_k - {f^{jk}}_i s\beta_j \bar \beta_k^{(n)} \, .   
\label{poiss_e34}  
\end{eqnarray}  
Therefore  
\begin{eqnarray}  
\Delta {\cal A}^{(n)}_i & = &   
s {\cal A}^{(n)}_i - {f^{jk}}_i {\cal A}^{(n)}_k \beta_j \nonumber \\  
& = & -{f^{jk}}_i s \bar \beta_j^{(n)} \beta_k - {f^{jk}}_i s \beta_j   
\bar \beta^{(n)}_k \nonumber \\  
&  & -{f^{jk}}_i \left ( s \bar \beta_k^{(n)}  - {f^{pq}}_k
\bar \beta^{(n)}_p \beta_q \right ) \beta_j \nonumber \\  
& = & 0   
\label{poiss_e35}  
\end{eqnarray}  
by the Jacobi identity.  

\section{Superfield action of the BV differential in the gauge sector}
\label{appB}

We wish to compute the $\delta$-variation of $\Sf{\eta}_i$
given in eq.(\ref{pr20}).

As a preliminary step we evaluate
the derivative w.r.t. $\Sf{X}^r$ of eq.(\ref{pr21}).
The result is 
\begin{eqnarray}
\frac{\partial}{\partial \Sf{X}^r} \alpha^{ij}(\Sf{\bar X}(\Sf{X})) & = & 
\left . \frac{\partial \alpha^{ij}}{\partial \Sf{\bar X}^l} \right |_{\Sf{\bar X}=\Sf{\bar X}(\Sf{X})}
\frac{\partial \Sf{\bar X}^l}{\partial \Sf{X}^r} \nonumber \\
& = & \frac{\partial^2 \Sf{\bar X}^i}{\partial \Sf{X}^r \partial \Sf{X}^p}
      \frac{\partial \Sf{\bar X}^j}{\partial \Sf{X}^q} {f^{pq}}_v \Sf{X}^v 
     +\frac{\partial \Sf{\bar X}^i}{\partial \Sf{X}^p} 
      \frac{\partial^2 \Sf{\bar X}^j}{\partial \Sf{X}^r \partial \Sf{X}^q}
      {f^{pq}}_v \Sf{X}^v \nonumber \\
&   & + \frac{\partial \Sf{\bar X}^i}{\partial \Sf{X}^p} \frac{\partial \Sf{\bar X}^j}{\partial \Sf{X}^q} {f^{pq}}_r \, ,
\label{pr23}
\end{eqnarray}
from which one gets
\begin{eqnarray}
\left . \frac{\partial \alpha^{ij}}{\partial \Sf{\bar X}^l} \right |_{\Sf{\bar X}=\Sf{\bar X}(\Sf{X})} & = &
\frac{\partial \Sf{X}^r}{\partial \Sf{\bar X}^l} \Big [
\frac{\partial^2 \Sf{\bar X}^i}{\partial \Sf{X}^r \partial \Sf{X}^p}
      \frac{\partial \Sf{\bar X}^j}{\partial \Sf{X}^q} {f^{pq}}_v \Sf{X}^v 
     +\frac{\partial \Sf{\bar X}^i}{\partial \Sf{X}^p} 
      \frac{\partial^2 \Sf{\bar X}^j}{\partial \Sf{X}^r \partial \Sf{X}^q}
      {f^{pq}}_v \Sf{X}^v \Big ] \nonumber \\
& & + \frac{\partial \Sf{\bar X}^i}{\partial \Sf{X}^p} \frac{\partial \Sf{\bar X}^j}{\partial \Sf{X}^q} {f^{pq}}_r \frac{\partial \Sf{X}^r}{\partial \Sf{X}^l} \, .
\label{pr24}
\end{eqnarray}
Now we compute the $\delta$-variation of $\Sf{\eta}_i$:
\begin{eqnarray}
\delta \Sf{\eta}_i & = & 
\frac{\partial^2 \Sf{\bar X}^j}{\partial \Sf{X}^p \partial \Sf{X}^i} \delta \Sf{X}^p \Sf{\bar \eta}_j + \frac{\partial \Sf{\bar X}^j}{\partial \Sf{X}^i} \delta \Sf{\bar \eta}_j \nonumber \\
& = & \frac{\partial^2 \Sf{\bar X}^j}{\partial \Sf{X}^p \partial \Sf{X}^i}
      ( D \Sf{X}^p + \alpha^{pq}(\Sf{X}) \Sf{\eta}_q) \Sf{\bar \eta}_j
      + \frac{\partial \Sf{\bar X}^j}{\partial \Sf{X}^i}
      ( D \Sf{\bar \eta}_j + \frac{1}{2} \partial_j \alpha^{pq} \Sf{\bar \eta}_p \Sf{\bar \eta}_q )
\nonumber \\
& = & \frac{\partial^2 \Sf{\bar X}^j}{\partial \Sf{X}^p \partial \Sf{X}^i} D \Sf{X}^p \Sf{\bar \eta}_j
      + \frac{\partial \Sf{\bar X}^j}{\partial \Sf{X}^i} D \Sf{\bar \eta}_j \nonumber \\
&   & + \frac{\partial^2 \Sf{\bar X}^j}{\partial \Sf{X}^p \partial \Sf{X}^i}
        \alpha^{pq}(\Sf{X}) \Sf{\eta}_q \Sf{\bar \eta}_j
      + \frac{1}{2} \frac{\partial \Sf{\bar X}^j}{\partial \Sf{X}^i} \partial_j \alpha^{pq} \Sf{\bar \eta}_p \Sf{\bar \eta}_q 
\nonumber \\
& = & D  \Sf{\eta}_i \nonumber \\
&   & + \frac{\partial^2 \Sf{\bar X}^j}{\partial \Sf{X}^p \partial \Sf{X}^i}
        \alpha^{pq}(\Sf{X}) \Sf{\eta}_q \Sf{\bar \eta}_j
      + \frac{1}{2} \frac{\partial \Sf{\bar X}^j}{\partial \Sf{X}^i} \partial_j \alpha^{pq} \Sf{\bar \eta}_p \Sf{\bar\eta}_q \, .
\label{pr25}
\end{eqnarray}
We denote
\begin{eqnarray}
&& A_i = \frac{1}{2} \frac{\partial \Sf{\bar X}^j}{\partial \Sf{X}^i} \partial_j \alpha^{pq} \Sf{\bar \eta}_p \Sf{\bar \eta}_q \, ,
\nonumber \\
&& B_i = \frac{\partial^2 \Sf{\bar X}^j}{\partial \Sf{X}^p \partial \Sf{X}^i}
        \alpha^{pq}(\Sf{X}) \Sf{\eta}_q \Sf{\bar \eta}_j \, .
\label{pr26}
\end{eqnarray}
We insert eq.(\ref{pr24}) into $A_i$ in the above equation and get
\begin{eqnarray}
A_i & = & \frac{\partial \Sf{\bar X}^j}{\partial \Sf{X}^i} \Big [ \frac{1}{2}
    \frac{\partial \Sf{X}^r}{\partial \Sf{\bar X}^j} \Big [
\frac{\partial^2 \Sf{\bar X}^p}{\partial \Sf{X}^r \partial \Sf{X}^a}
      \frac{\partial \Sf{\bar X}^q}{\partial \Sf{X}^b} {f^{ab}}_v \Sf{X}^v 
     +\frac{\partial \Sf{\bar X}^p}{\partial \Sf{X}^a} 
      \frac{\partial^2 \Sf{\bar X}^q}{\partial \Sf{X}^r \partial \Sf{X}^b}
      {f^{ab}}_v \Sf{X}^v \Big ] \nonumber \\
& & + \frac{1}{2} \frac{\partial \Sf{\bar X}^p}{\partial \Sf{X}^a} \frac{\partial \Sf{\bar X}^q}{\partial \Sf{X}^b} {f^{ab}}_r \frac{\partial \Sf{X}^r}{\partial \Sf{\bar X}^j} \Big ] \Sf{\bar \eta}_p \Sf{\bar \eta}_q \nonumber \\    
& = & \frac{1}{2} \Big ( \frac{\partial^2 \Sf{\bar X}^p}{\partial \Sf{X}^i \partial \Sf{X}^a}
                       \frac{\partial\Sf{\bar X}^q}{\partial \Sf{X}^b} {f^{ab}}_v \Sf{X}^v
                      + \frac{\partial \Sf{\bar X}^p}{\partial \Sf{X}^a} 
                        \frac{\partial^2 \Sf{\bar X}^q}{\partial \Sf{X}^i \partial \Sf{X}^b}
			{f^{ab}}_v \Sf{X}^v \Big ) \Sf{\bar \eta}_p \Sf{\bar \eta}_q \nonumber \\ 
& & + \frac{1}{2} \frac{\partial \Sf{\bar X}^p}{\partial \Sf{X}^a} \frac{\partial \Sf{\bar X}^q}{\partial \Sf{X}^b} 
{f^{ab}}_i \Sf{\bar \eta}_p \Sf{\bar \eta}_q \, .
\label{pr27}
\end{eqnarray}
Moreover
\begin{eqnarray}
B_i = \frac{\partial^2 \Sf{\bar X}^j}{\partial \Sf{X}^p \partial \Sf{X}^i}
        {f^{pq}}_v \Sf{X}^v \Sf{\eta}_q \Sf{\bar \eta}_j \, .
\label{pr28}
\end{eqnarray}
Now one can compute the sum $A_i + B_i$. By taking into account the fact that
the $\Sf{\bar \eta}$'s are odd the result is
\begin{eqnarray}
A_i + B_i = \frac{1}{2} \frac{\partial \Sf{\bar X}^p}{\partial \Sf{X}^a} \frac{\partial \Sf{\bar X}^q}{\partial \Sf{X}^b}
   {f^{ab}}_i \Sf{\bar \eta}_p \Sf{\bar \eta}_q = \frac{1}{2} {f^{ab}}_i \Sf{\eta}_a \Sf{\eta}_b \, .
\label{pr29}
\end{eqnarray}
By substituting back into eq.(\ref{pr25}) one finally finds
\begin{eqnarray}
\delta \Sf{\eta}_i = D \Sf{\eta}_i +  \frac{1}{2} {f^{ab}}_i \Sf{\eta}_a \Sf{\eta}_b \, ,
\label{pr30}
\end{eqnarray}
which corresponds to the action of $\delta$ on the superfield $\Sf{\eta}_i$ in presence of a linearized
Poisson tensor.


\end{document}